\documentclass[cameraready]{Interspeech}

\title{Automated Measurement of Geniohyoid Muscle Thickness During Speech Using Deep Learning and Ultrasound}

\author[affiliation={1}, orcid=0000-0003-4460-4204]{Alisher}{Myrgyyassov}
\author[affiliation={2, 4}, orcid=0000-0003-3564-2911]{Bruce Xiao}{Wang}
\author[affiliation={1}, orcid=0000-0002-4539-8089]{Yu}{Sun}
\author[affiliation={2}, orcid=0009-0005-6053-6063]{Shuming}{Huang}
\author[affiliation={1}, orcid=0000-0002-2035-2420]{Zhen}{Song}
\author[affiliation={3, 4}, orcid=0000-0003-0585-3566]{Min Ney}{Wong}
\author[affiliation={1, 4}, orcid=0000-0002-3407-9226, correspondingauthor]{Yongping}{Zheng}

\address{
    $^1$ Biomedical Engineering Department, Hong Kong Polytechnic University, Hong Kong, China \\
    $^2$ English and Communication Department, Hong Kong Polytechnic University, Hong Kong, China \\
    $^3$ Department of Language Science and Technology, Hong Kong Polytechnic University, Hong Kong, China
    $^4$ Research Institute for Smart Ageing, Hong Kong Polytechnic University, Hong Kong, China
}

\email{alisher2.myrgyyassov@connect.polyu.hk, brucex.wang@polyu.edu.hk, stefanie.sun@polyu.edu.hk, shu-ming.huang@connect.polyu.hk, zhen0212.song@connect.polyu.hk, min.wong@polyu.edu.hk, yongping.zheng@polyu.edu.hk}

\keywords{ultrasound imaging, geniohyoid muscle, articulatory phonetics, deep learning, automated measurement}

\usepackage{comment}
\usepackage{tipa}

\newcommand{\along}{ /a\textlengthmark/ }
\newcommand{\ilong}{ /i\textlengthmark/ }
\newcommand{\ulong}{ /u\textlengthmark/ }

\begin{document}

\maketitle

\begin{abstract}
    Manual measurement of muscle morphology from ultrasound during speech is time-consuming and limits large-scale studies. We present SMMA, a fully automated framework combining deep learning segmentation with skeleton-based thickness quantification for analyzing geniohyoid (GH) muscle dynamics. Validation demonstrates near-human-level accuracy (Dice: 0.9037, MAE: 0.53 mm, r = 0.901). Application to Cantonese vowel production (N = 11) reveals systematic patterns: \along shows significantly greater GH thickness (7.29 mm) than \ilong (5.95 mm, \(p < 0.001\), Cohen’s \(d > 1.3\)), suggesting greater GH activation during the production of \along than \ilong, consistent with its role in mandibular depression. Sex differences (5-8\% greater in males) reflect anatomical scaling. SMMA achieves expert-validated accuracy while eliminating the need for manual annotation, enabling scalable investigations of speech motor control and objective assessment of speech/swallowing disorders.
\end{abstract}

\section{Introduction}
{\let\thefootnote\relax\footnote{Submitted to Interspeech 2026}}
Ultrasound imaging has emerged as a valuable tool for studying tongue dynamics during speech production, offering real-time visualization of articulatory movements without radiation exposure or positional constraints. However, ultrasound-based articulatory analysis has predominantly focused on tongue contour tracking \cite{al2022tongue}, which benefits from relatively straightforward automated segmentation methods and provides rich phonetic information about tongue shape configurations during speech. In contrast, the deeper geniohyoid (GH) tongue muscle remains largely unexplored in speech research despite its critical role in influencing tongue body position and overall oral cavity configuration via its effects on hyoid and jaw mechanics \cite{khan2025anatomy}. Because tongue body position is central to the realization of vowel height and backness, as well as certain consonantal constrictions, GH activity is expected to have direct consequences for articulatory patterns in speech \cite{matsuo2010kinematic}.

This gap arises from inherent technical and anatomical challenges: these muscles have historically been difficult to visualize clearly in ultrasound images \cite{vander2024feasibility}, and the measurements are influenced by probe placement \cite{pauloski2025reliability} and variability between testers due to differences in interpretation and annotation \cite{shimizu2016retest}. Recent advances in ultrasound imaging systems have substantially improved muscle visibility, creating new opportunities for systematic investigation. However, progress remains constrained by the lack of validated automated segmentation tools, as manual delineation of muscle boundaries is time-consuming, subjective, and susceptible to inter-rater variability, creating a significant bottleneck for large-scale phonetic and clinical studies.

Hence, the GH muscle, despite its anatomical significance in tongue positioning and articulation, has received limited attention in the phonetic literature due to imaging and methodological challenges. Existing studies that have attempted to measure tongue muscle properties from ultrasound images have relied on manual measurements \cite{watkin2001ultrasonic, kelly2024suprahyoid, shimizu2016retest} with considerable inter-rater variability and have focused primarily on physiology, such as eating and swallowing assessment \cite{kelly2024suprahyoid, maeda2023ultrasonography}, as well as muscle thickness in sarcopenia patients \cite{barotsis2020muscle-thickness}, rather than articulation. 

In this work, we propose a standardized and comprehensive methodology for fully-automatic measurement of the GH muscle on B-mode ultrasound mid-sagittal images, powered by a deep learning-based segmentation model. Our approach eliminates the need for manual annotation by automating the process of segmenting the GH muscle, followed by skeletonization and distance measurement from the skeleton to the edges of the derived segmentation mask. To validate our method, we assess segmentation quality through inter-annotator agreement (three annotators trained by an experienced sonographer with 10 years of clinical practice) and measure thickness accuracy against manual measurements performed by the sonographer. By introducing and validating this algorithm, we aim to enhance the accessibility and repeatability of muscle measurement for further physiological, anatomical, and speech research.

\section{Methodology}
\subsection{Overview}
This study introduces and validates SMMA (Skeleton-based Morphometric Muscle Analysis), a novel computational framework for extracting kinematic parameters of the geniohyoid muscle from ultrasound video during speech production. The SMMA framework integrates two core components: (1) deep learning-based automated segmentation of muscle boundaries, and (2) skeleton-based extraction of muscle thickness, with temporal alignment to phonetic events. Hence, we perform two separate validation experiments to assess the performance of each component using our own ultrasound dataset. 

\subsection{Component 1: Segmentation}
Component 1 performs image pre-processing and segmentation. Ultrasound frames undergo standardization (cropping, resizing, normalization) before being processed by a convolutional neural network trained to delineate GH muscle boundaries. The output is a binary segmentation mask representing the muscle region of interest. Both the input image and the binary mask are 224×224px.

A segmentation deep learning model is a core part of Component 1 that extracts the area of the GH muscle automatically without needing initialization. To ensure fair comparison across all models evaluated in this study, we standardize the training protocol as follows: each model is trained for 50 epochs with early stopping applied after 10 epochs without improvement. We employed a combined loss function using Dice \cite{dice} and Focal \cite{focal} losses with weights of 0.8 and 0.2, respectively. Ultrasound-specific online augmentations are employed as described in \cite{usaugment}. The learning rate and initialization for each model are selected based on the respective original training settings reported in the literature.

\subsection{Component 2: Thickness Extraction}
Component 2 applies morphometric analysis to the segmentation mask. A skeletonization algorithm extracts the muscle's medial axis, from which thickness (perpendicular distance between boundaries at specified points along the skeleton) and length (skeleton arc length) are computed. These measurements can be extracted from individual frames or across temporal sequences, enabling both static morphological characterization and dynamic articulatory analysis. The segmentation masks are post-processed using a combination of morphological operations (closing, opening, and hole filling) to remove small artifacts and Gaussian smoothing to smooth edges. Additionally, the algorithm preserves connectivity by ensuring the largest connected component is retained and prevents splitting during the morphological and smoothing processes.

After post-processing, the masks are skeletonized using the skeletonization algorithm \cite{skelet}, producing the one-pixel wide "spine":
\begin{align}
S &= \text{Skel}(M_{\text{smooth}})
\end{align}
Where \(M_{\text{smooth}}\) is the post-processed segmentation mask by Gaussian smoothing and \(S\) denotes all skeleton points.

For each skeleton point \( p \), the local thickness $t(p)$ is calculated as double the distance \( d(p) \) from \( p \) to the opposite sides of the muscle boundary \( \partial M \). To reduce sensitivity to edge effects, only the interquartile range (25\%-75\%) of the skeleton points is used to compute the mean thickness:
\begin{align}
T_{\text{mean}} &= \frac{1}{|S_{\text{middle}}|} \sum_{p \in S_{\text{middle}}} t(p)
\end{align}
where \( S_{\text{middle}} \) denotes the skeleton points in the 25th to 75th percentile.

\begin{figure} 
    \centering
    \includegraphics[width=0.95\linewidth]{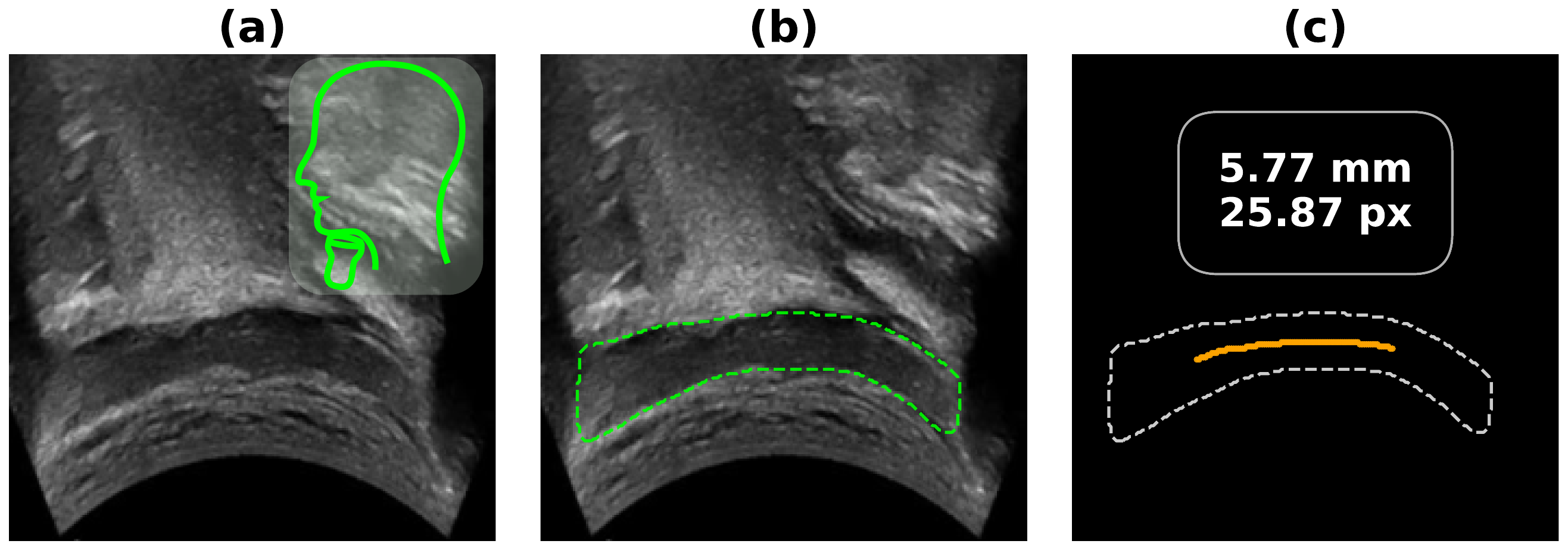} 
    \caption{Visualization of the SMMA pipeline. Image (a) shows the original image and probe placement, (b) an automatically generated mask by UltraUNet, and (c) shows the middle 50\% skeleton extracted from the mask and corresponding thickness measurements in pixels (px) and millimetres (mm).}
    \label{fig:smma-pipeline}
\end{figure}

Optionally, the user can get the skeleton curvature \cite{coeurjolly2001discrete} and cross-sectional area of the mask for morphometric analysis. The two-component pipeline is visualized in Figure \ref{fig:smma-pipeline}.

\subsection{Datasets and Data Collection Protocols}
The dataset consists of 1650 annotated ultrasound images from 5 male and 6 female Cantonese speakers. From each subject's recording session (tasks include: isolated vowels, consonants, syllable sequences), 150 images were sampled at a consistent rate. All sampled images were annotated, with approximately 12\% of images with suboptimal quality (blurry edges, acoustic shadows), to train the model on realistic clinical data variability. Data were split 7:2:2 (subjects) into train/validation/test subsets. In Component 1 validation, three trained annotators manually labeled the ultrasound images. To assess inter‑rater reliability, each image in the test subset was independently annotated by all three annotators, while images in the training and validation sets received a single annotation each.

Ultrasound high-resolution B-mode imaging data were acquired using a SuperSonic Imagine Aixplorer ultrasound scanner equipped with an SC6-1 convex probe (Aix-en-Provence, France). During the data collection process, participants were seated comfortably in an upright position to ensure consistent imaging conditions and natural tongue movement. A trained sonographer positioned the probe along the midsagittal plane beneath the chin, between the hyoid and mandible bones, to visualize the GH muscle with the intermediate tendon as an anatomical landmark to ensure consistent imaging plane across subjects and sessions. The same sonographer performed all data collection to minimize inter-operator variability. Ultrasound video was recorded at 30 frames per second, with a synchronized audio recording.

\subsection{Validation}
The validation of SMMA assesses each of the proposed components separately. Component 1 validation compares the segmentation accuracy of deep learning models with the annotations of trained annotators. Selected models include Attention UNet \cite{attunet}, UNet \cite{attunet}, UltraUNet \cite{myrgyyassov}, SwinUNet \cite{swin}, and Deeplab v3 \cite{deeplabv3} with ResNet-50 \cite{resnet} as the backbone, as they are successfully applied and validated in the domain of ultrasound tongue imaging segmentation on similar datasets \cite{myrgyyassov}. To assess the segmentation quality of the GH muscle boundaries obtained from the models, we compare their results on the test dataset taken from the unseen subjects' data and annotated by three different experts independently. In the test dataset, each image was annotated three times, and the expert annotations are also compared with each other to assess the variability between the manual annotations. Each model was trained independently three times during Component 1 validation, and the averaged performance metrics were reported.

To validate the thickness measurement capabilities (Component 2) of the proposed algorithm, the sonographer manually measured the thickness of the GH muscles in 110 images (10 images per subject) unseen by the segmentation models. Of these, 55 images are randomly selected without considering image quality, while the other 55 were chosen by the sonographer for their clinical importance. The clinically important images are typically selected from the temporal middle point based on acoustic signals within the imaging sequence for each isolated vowel, where the muscles are clearly visible. 

\subsection{Applications}
To demonstrate the possible clinical applications of the designed methodology in articulatory analysis, we dynamically measure the GH muscle thickness during the pronunciation of isolated vowels \along, \ilong, and \ulong for each subject and compare the results with the existing knowledge about suprahyoid muscles involvement during speech production. Additionally, we compare the muscle thicknesses between sexes. Each vowel was pronounced six times during the speech task, making a total of 66 utterances for each vowel.

\section{Results}

\subsection{Component 1 Validation}

This evaluation stage assesses the segmentation component of SMMA by assessing the reliability of expert annotations and benchmarking candidate models against this reference.

\begin{table}[t]
  \centering
  \caption{Component 1 validation summary: human inter-annotator agreement and model performance (averaged across subjects and runs, with $\pm$ indicating run-wise standard deviation).}
  \label{tab:stage1}
  \begin{tabular}{lccc}
    \toprule
    Pair / Model & Dice (Mean~$\pm$~SD) & IoU (Mean~$\pm$~SD) \\
    \midrule
    Ann1 vs Ann2 & $0.9179 \pm 0.0247$ & $0.8493 \pm 0.0410$ \\
    Ann1 vs Ann3 & $0.9036 \pm 0.0444$ & $0.8270 \pm 0.0682$ \\
    Ann2 vs Ann3 & $0.9001 \pm 0.0423$ & $0.8209 \pm 0.0654$ \\
    \midrule
    Attention UNet & $0.8707 \pm 0.0332$ & $0.7801 \pm 0.0483$ \\
    DeepLab v3    & $0.8792 \pm 0.0175$ & $0.7874 \pm 0.0269$ \\
    SwinUNet      & $0.8159 \pm 0.0746$ & $0.7006 \pm 0.1046$ \\
    UltraUNet     & $\mathbf{0.9037} \pm \mathbf{0.0035}$ & $\mathbf{0.8263} \pm \mathbf{0.0057}$\\
    UNet          & $0.8870 \pm 0.0193$ & $0.8016 \pm 0.0298$ \\
    \bottomrule
  \end{tabular}
\end{table}

Inter-annotator agreement on the test set is high, indicating consistent manual delineation of GH boundaries. Across all subjects, annotator pairs achieve mean Dice ranging from $0.9001 \pm 0.0423$ to $0.9179 \pm 0.0247$ and IoU from $0.8209 \pm 0.0654$ to $0.8493 \pm 0.0410$. These values define the human performance range for subsequent model comparison.

We then compare five segmentation architectures on the same test images. The most relevant overall metrics (averaged across three runs and subjects) are summarised in Table~\ref{tab:stage1}. UltraUNet achieves the best trade-off between accuracy and stability, with mean Dice and IoU close to inter-annotator agreement and the lowest variability across runs (image-wise Dice SD $0.0086$, IoU SD $0.0138$; run-wise Dice SD $0.0318$, IoU SD $0.0479$). SwinUNet shows clearly inferior performance, while UNet and DeepLab v3 perform competitively but below UltraUNet. Furthermore, UltraUNet (4.45M parameters, 250 masks/s, GPU RTX3060, CPU 11th Gen Intel Core i5-11400) achieves a run-wise HD95 of $2.25 \pm 0.32$, outperforming the second-best and second-fastest UNet (31.04M params, 90 masks/s) in both accuracy and speed, which achieves $3.54 \pm 1.47$.

Given its accuracy close to the human range, low HD95, and minimal across-run variability, UltraUNet is selected as the primary segmentation backbone for SMMA and is used in all subsequent Component 2 and vowel analyses.

\subsection{Component 2 Validation}

This stage evaluates the thickness measurement component of SMMA by comparing automated skeleton-based measurements against manual ground truth annotations performed by an experienced clinical sonographer.

\begin{figure} 
    \centering
    \includegraphics[width=0.95\linewidth]{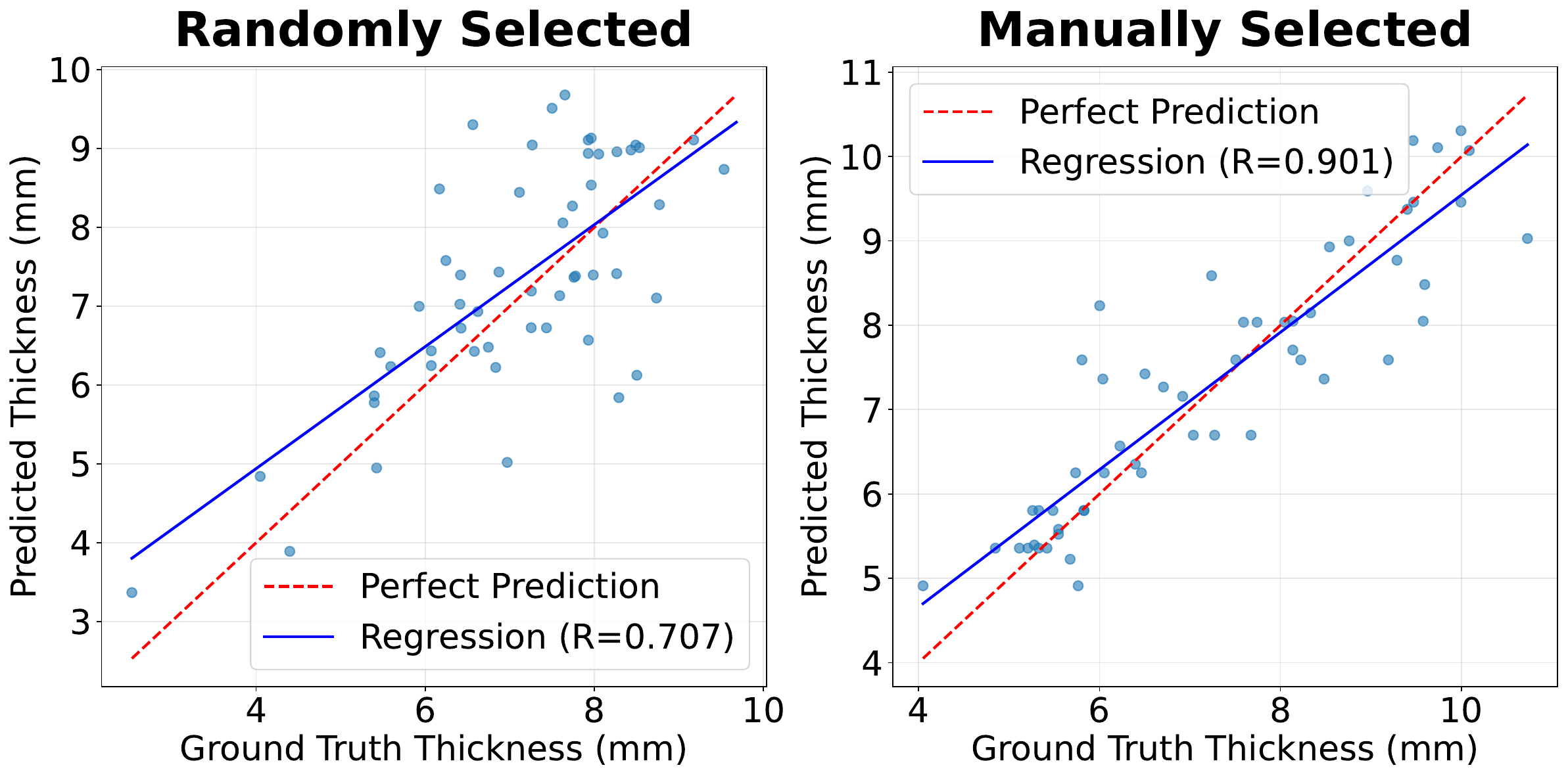} 
    \caption{Component 2 thickness measurement validation: SMMA automated measurements (left) vs. sonographer ground truth annotations (right).}
    \label{fig:linear-graphs}
\end{figure}

\begin{figure*}[tp] 
    \centering
    \includegraphics[width=0.9\textwidth]{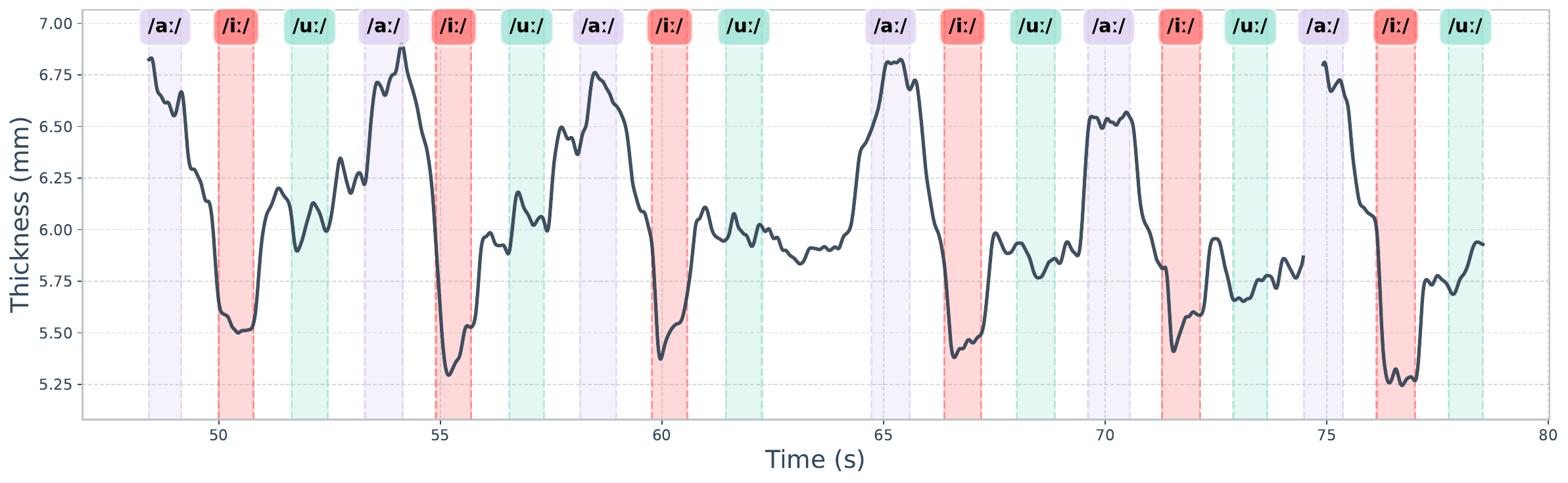} 
    \caption{Representative 30-second sample from the full recording of continuous muscle thickness tracking on a female subject during speech production. Subject produces repeated \along-\ilong-\ulong isolated vowel sequences  (purple = \along, red = \ilong, green = \ulong, white = pause), each vowel lasting around 800 ms. Brief gaps in tracking occur when image quality temporarily degrades (e.g., ~75s), demonstrating algorithm behavior under naturalistic recording conditions.}
    \label{fig:thickness}
\end{figure*}

Figure \ref{fig:linear-graphs} summarizes the thickness measurement validation results. For randomly selected images, SMMA achieves a mean absolute error (MAE) of 0.88 mm and RMSE of 1.10 mm, with strong correlation to ground truth (Pearson r = 0.707, $p < 0.001$, $R^2$ = 0.317). The mean relative error is 12.75\%, with 95\% of measurements falling within $\pm 1.83$ mm of manual annotations (Bland-Altman limits of agreement). For clinically selected images with superior visualization, performance improves substantially: MAE = 0.53 mm, RMSE = 0.75 mm, with very strong correlation (r = 0.901, $p < 0.001$, $R^2$ = 0.810) and mean relative error of 7.57\%. The 95\% limits of agreement narrow to $\pm 1.46$ mm, indicating excellent clinical agreement.

Linear regression analysis reveals that SMMA measurements scale predictably with ground truth values (randomly selected: slope = 0.77, intercept = 1.84 mm; clinically selected: slope = 0.81, intercept = 1.40 mm). Both regression lines approach the identity line (perfect prediction) more closely in high-quality images (Figure \ref{fig:linear-graphs}), with minimal systematic bias (Bland-Altman mean difference: 0.24 mm for random, 0.06 mm for clinical images).

\subsection{Application in Isolated Vowel Analysis}

To demonstrate the applicability of our measurement approach, we analyzed the geniohyoid muscle thickness during isolated vowel production in a cohort of 11 healthy Cantonese-speaking adults (6 female, 5 male, age: 24--35 years). Participants produced three vowels (\along, \ilong, \ulong) with six repetitions each, resulting in 198 measurements. The measurement protocol demonstrated good repeatability with an overall coefficient of variation of 5.06\% (SD: 3.74\%).

Significant differences in muscle thickness were observed across vowels (repeated measures ANOVA: $F = 4.84$, $p = 0.019$), with \along showing the greatest thickness (7.29 $\pm$ 0.90 mm), followed by \ulong (6.65 $\pm$ 0.79 mm) and \ilong (5.95 $\pm$ 0.84 mm). The comparison between \along and \ilong yielded large effect sizes in both females (Cohen's $d = 1.86$, $p < 0.001$) and males (Cohen's $d = 1.31$, $p < 0.001$). These findings align with established articulatory phonetics, as \along requires greater mouth opening and jaw lowering compared to the high front vowel \ilong \cite{scharinger2016linguistic, kawahara2014quantifying}, resulting in increased suprahyoid muscle activation \cite{hara2018treatment, kajisa2018relationship}.


Sex-based differences in absolute thickness were observed for \ilong ($p = 0.041$) and \ulong ($p = 0.009$) but not for \along ($p = 0.200$), with males showing 5--8\% greater thickness. A linear mixed-effects model with vowel type and sex as fixed effects and random intercepts by subject (to account for repeated measurements) confirmed the significant effect of vowel type ($p < 0.001$) while the vowel-by-sex interaction was not significant, suggesting that relative muscle thickness patterns across vowels are consistent between sexes. Figure~\ref{fig:thickness} illustrates representative thickness measurements across the three vowel conditions.

\begin{table}[h]
\centering
\caption{Geniohyoid muscle thickness across vowels and sex}
\label{tab:vowel_thickness}

\begin{tabular}{lccc}
\hline
\textbf{Vowel} & \textbf{Female (mm)} & \textbf{Male (mm)} & \textbf{\(p\)-value} \\
\hline
\along & \(7.16 \pm 0.73\) & \(7.45 \pm 1.05\) & 0.200 \\
\ilong  & \(5.76 \pm 0.77\) & \(6.18 \pm 0.87\) & 0.041* \\
\ulong & \(6.41 \pm 0.68\) & \(6.92 \pm 0.84\) & 0.009** \\
\hline
\multicolumn{4}{l}{\footnotesize *\(p < 0.05\), **\(p < 0.01\)} \\
\end{tabular}

\end{table}

\section{Discussion}

This study introduces SMMA, a fully automated framework for quantifying geniohyoid muscle morphology from ultrasound video during speech production. Two-stage validation demonstrates that UltraUNet achieves near-human-level segmentation accuracy (Dice: 0.9037), while skeleton-based thickness measurements show strong agreement with expert annotations (MAE: 0.53 mm, r = 0.901). Application to isolated vowel production reveals systematic patterns: \along produces significantly higher GH thickness (7.29 mm) than \ilong (5.95 mm), consistent with increased muscle activation during jaw lowering for low vowels \cite{hara2018treatment}.

The observed thickness differences likely reflect activation level rather than passive stretching \cite{han2013automatic, stokes2021methodological}. When muscles contract, they typically shorten and thicken due to volume conservation \cite{wakeling2014transverse, almohimeed2020ultrasound}. The GH muscle's primary function—mandibular depression—suggests that \along production requires active contraction to lower the jaw, increasing thickness, while \ilong production involves jaw raising with reduced GH activation, explaining thinner measurements. Large effect sizes between \along and \ilong (Cohen's $d > 1.3$) indicate robust activation patterns.

Sex-based differences in absolute thickness (5-8\% greater in males) likely reflect anatomical size rather than functional differences. Future work should normalize measurements by anatomical landmarks to isolate physiological from structural variation.

The skeleton-based algorithm shows strong clinical agreement but assumes relatively uniform morphology. As a result, irregular configurations may introduce artifacts. Validation focused on isolated vowels with straightforward temporal segmentation; extending to continuous speech requires robust phonetic alignment to handle coarticulation and rapid transitions. Automated GH measurement could potentially facilitate large-scale studies of speech motor control, enable objective dysarthria assessment \cite{mori2024cutoff}, and support rehabilitation monitoring.

Limitations include a relatively modest sample size (N=11), single language (Cantonese), and reliance on one sonographer for thickness measurement ground truth. Image quality substantially affects accuracy (MAE: 0.53 mm for high-quality vs. 0.88 mm for random images). Future work should establish quality thresholds, validate across linguistic contexts and pathological populations, and assess longitudinal stability.

\section{Conclusion}
We present SMMA, a validated framework enabling fully automated, frame-by-frame measurement of geniohyoid muscle thickness from ultrasound video during speech. The method achieves clinical-grade accuracy ($MAE < 0.6 mm$) and reveals systematic articulatory patterns across vowels consistent with speech motor physiology. By eliminating manual annotation bottlenecks, SMMA enhances accessibility and reproducibility of muscle-based articulatory research, with applications in speech science, clinical assessment, and rehabilitation monitoring.



\section{Acknowledgements}
This project is funded by the Young Innovative Researcher Award Scheme (P0057955) of The Hong Kong Polytechnic University. This work involved human subjects in its research, and approval of all ethical and experimental procedures and protocols was granted by the Human Subjects Ethics Sub-Committee of The Hong Kong Polytechnic University under application No. HSEARS20240306010. The work was supported in part by the Research Grants Council of Hong Kong under Grant GRF No. 15217224.

\section{Generative AI Use Disclosure}
Generative AI tools (e.g., large language models) were used to assist with language editing and rewriting for clarity, including grammar correction, rephrasing, and summarization of text originally written by the authors. They were not used to generate new scientific ideas, design experiments, perform data analysis, or write significant original portions of the manuscript. All conceptual contributions, experimental designs, and interpretations of results are entirely due to the human authors, who take full responsibility for the content.


\bibliographystyle{IEEEtran}
\bibliography{mybib}

\end{document}